\documentclass[aps, preprint, onecolumn,secnumarabic,nobalancelastpage,nofootinbib]{revtex4-1} 

\usepackage{natbib}
\usepackage{graphicx}
\usepackage{amsmath,amssymb}
\usepackage[utf8]{inputenc}
\usepackage[T1]{fontenc}
\usepackage{lmodern}
\usepackage{hyperref}
\usepackage{xcolor}
\hypersetup{colorlinks=true,citecolor={blue},linkcolor={blue},urlcolor={blue}}
\usepackage{units}
\usepackage[english]{babel}
\usepackage[normalem]{ulem}
\usepackage{upgreek}
\usepackage{wasysym}
\usepackage{placeins} 
\usepackage{soul}
\makeindex

\begin{document}
	
\title{Synthetic band-structure engineering in polariton crystals with non-Hermitian topological phases}

\date{\today}

\author{L. Pickup$^{1}$}
\author{H. Sigurdsson$^{1,2}$}
\author{J. Ruostekoski$^{3}$}
\author{P. G. Lagoudakis$^{1,2*}$}
\affiliation{$^1$Department of Physics and  Astronomy, University of Southampton, Southampton, SO17 1BJ, United Kingdom}
\affiliation{$^2$Skolkovo Institute of Science and Technology Novaya St., 100, Skolkovo 143025, Russian Federation}
\affiliation{$^3$Physics Department, Lancaster University, Lancaster LA1 4YB, United Kingdom}
\affiliation{$^*$email: Lucy.Pickup@soton.ac.uk, Pavlos.Lagoudakis@soton.ac.uk}

\renewcommand{\abstractname}{} 
\begin{abstract}
	Synthetic crystal lattices provide ideal environments for simulating and exploring the band structure of solid-state materials in clean and controlled experimental settings. Physical realisations have, so far, dominantly focused on implementing irreversible patterning of the system, or interference techniques such as optical lattices of cold atoms. Here, we realise reprogrammable synthetic band-structure engineering in an all optical exciton-polariton lattice. We demonstrate polariton condensation into excited states of linear one-dimensional lattices, periodic rings, dimerised non-trivial topological phases, and defect modes utilising malleable optically imprinted non-Hermitian potential landscapes. The stable excited nature of the condensate lattice with strong interactions between sites results in an actively tuneable non-Hermitian analogue of the Su-Schrieffer-Heeger system.
\end{abstract}

\maketitle

\section*{Introduction}
Particles subjected to potential landscapes with discrete translational symmetries, whether natural or artificially made, exhibit bands of allowed energies corresponding to the quasimomentum of the crystal's Bloch states~\cite{ashcroft_solid}. For instance, electronic band theory explains the difference between insulating and conducting phases of materials, as well as their optical properties. With advances in energy band synthesis in atomic systems (optical lattices) or photonic crystals, complicated yet meticulous lattice investigations are now possible including superfluid-to-Mott insulator phase transitions~\cite{Greiner_Nature2002}, networks of Josephson junctions~\cite{Cataliotti_Science2001}, and solitonic excitations~\cite{Efremidis_PRE2002, Eiermann_PRL2004}. When the symmetry of a periodic structure is broken and/or boundaries are engineered in a desired way, there can arise defect states, surface states, and bound states in the continuum that do not dissipate energy into the surrounding environment. Advancements in photonics have allowed for the design and study of nearly lossless waveguides, filters and splitters~\cite{Joannopoulos_photonic_crystals}, with applications in communications and biomedicine. Recent developments have led to the study of topological states of matter in photonics~\cite{Ozawa_RevModPhys2019}, and separately in cold atoms ~\cite{Dalibard_RMP2011, Cooper_RevModPhys2019}.

One-dimensional (1D) crystals provide the simplest platform to study non-trivial topological phases, the prime example being the Su-Schrieffer-Heeger (SSH) model~\cite{Su_PRL1979}. Today, the Zak phase (or the 1D topological winding number) has been measured in a system of cold atoms~\cite{Atala_NatPhys2013}, followed by the demonstration of adiabatic Thouless pumping~\cite{Nakajima_NatPhys2016}, and an electronic topological superlattice~\cite{Belopolski_Science2017}. Recently, non-Hermitian solid-state and photonic systems have attracted a huge interest in the study of out-of-equilibrium topological phases~\cite{Xiao_PRX2014, Zeuner_PRL2015, Bandres_Science2018, Shen_PRL2018, Gong_PRX2018, Zhao_Science2019}, dissipative quantum physics~\cite{El-Ganainy2018, Zhong2018, El-Ganainy2019}, and the advantageous effects of unbroken parity-time symmetry~\cite{Regensburger_Nature2012}.
\begin{figure}
	\centering
	\includegraphics[width=15.6cm]{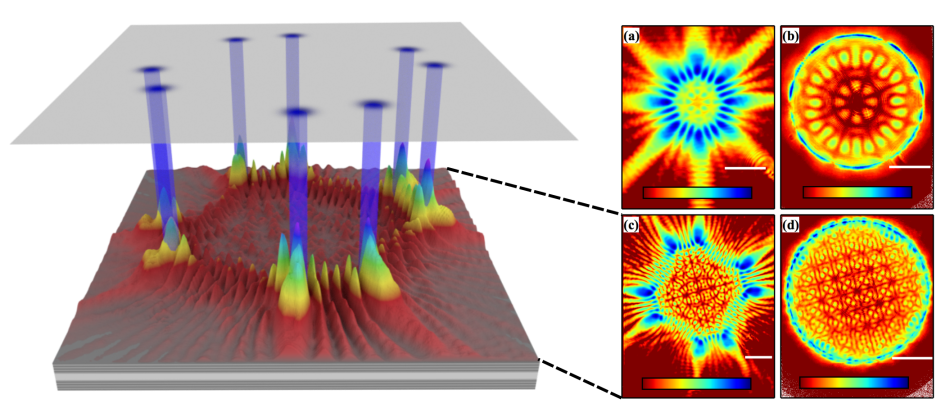}
	\caption{Experimental polariton condensate photoluminescence using eight pump spots forming a circle. Left panel shows the real space photoluminescence distribution, corresponding to (c), resulting from a staggered octagon of non-resonant optical beams, shown schematically by the blue cylinders. (a \& c) Real-space and (b \& d) $k$-space photoluminescence distributions using normalised logarithmic colour scales - shown at the bottom of each image. (a \& b) and (c \& d) show regular and dimarised octagons respectively. The white lines are $15$ $\upmu\mathrm{m}$ and  $1$ $\upmu\mathrm{m}^{-1}$ scale bars in (a \& c) and (b \& d) respectively.}
	\label{Fig_BlochBand_ExpOctagon}
\end{figure}

In the optical regime, a rapidly developing platform for the study of the above-mentioned phenomena are exciton-polaritons (from here on {\it polaritons}), realised in semiconductor microcavities. These hybrid light-matter quasi-particles are formed by the strong coupling of light confined in Fabry-P\'erot microcavities and electronic transitions in embedded semiconductor slabs~\cite{kavokin_microcavities_2011}. Their dissipative and out-of-equilibrium nature permits condensation into excited states~\cite{lai_coherent_2007, Maragkou_PRB2010, kim_dynamical_2011} that still presents a non-trivial task for cold atoms in thermal equilibrium~\cite{Wirth_NatPhys2010}.

In polaritonic systems there are two processes available to sculpt a crystal lattice. The most commonly applied process is through periodically patterning of the cavity mode and/or the intracavity quantum wells. This is typically achieved through patterned metallic deposition on top of the sample \cite{kim_dynamical_2011, lai_coherent_2007}, etch and overgrowth patterning techniques~\cite{Winkler_PRB2016}, surface acoustic waves~\cite{cerda-mendez_polariton_2010} or micro-structuring a sample into arrays of micropillars~\cite{StJean_NatPho2017, whittaker_exciton_2018, klembt_exciton-polariton_2018}. Linear features such as Dirac cones and flat bands have been demonstrated with polaritons utilising etched lattices in Lieb \cite{whittaker_exciton_2018} and honeycomb~\cite{jacqmin_direct_2014} geometries with topological transport recently reported~\cite{klembt_exciton-polariton_2018}, as well as non-linear dynamics of bright gap solitons~\cite{Cerda_PRL2013, Tanese_NatComm2013}. The other process utilises the matter component of polaritons to produce periodic potentials through many-body interactions. Similar to dipole moment induced optical traps for cold atoms~\cite{Bloch_NatPhys2005}, or photorefractive crystals~\cite{Neshev_PRL2004}, one can design an all-optical potential landscape for polaritons by using non-resonant optical excitation beams to create reservoirs of excitons which result in effective repulsive potentials due to polariton-exciton interactions~\cite{Wertz_NatPhys2010,tosi_sculpting_2012,tosi_geometrically_2012}.

In this article, we realise an all-optical, actively-tunable band-structure engineering platform harnessing reprogrammable non-Hermitian potential landscapes that result from interparticle interactions. Utilising this platform, we demonstrate a variety of band structure features including, polariton condensation into high-symmetry points in arbitrarily excited energy bands of the resulting Bloch states. By dimerising the potential landscape we experimentally realise an analogue of the topologically non-trivial SSH system, resulting in the formation of split energy band states. We determine through theoretical investigations that there is a $\pi$ change in the Zak phases of these bands when swapping the dimerisation of the potential landscape. This confirms that our system experimentally provides a platform for studying non-trivial topology in non-Hermitian systems. Finally by introducing local defects in the potentials periodicity, we demonstrate controllable highly localised defect state condensation with the resulting system containing analogues of bright and dark soliton modes.

\begin{figure}
	\centering
	\includegraphics[width=8.6cm]{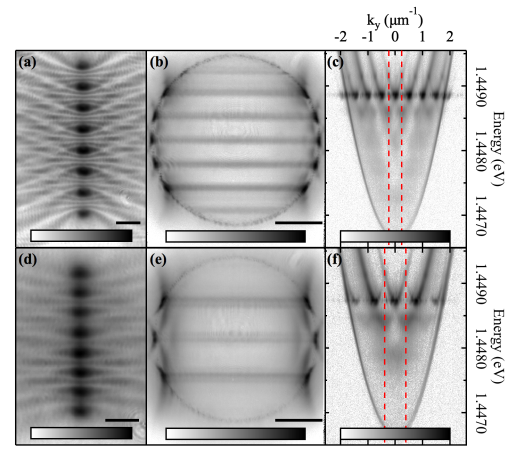}
	\caption{Experimental photoluminescence intensity distribution from a chain of eight equally separated polariton condensates in (a \& d) real-space and (b \& e) $k$-space and (c \& f) the corresponding dispersions. The lattice constant is $\approx 13$ $\upmu$m in (a,b,c) and $\approx 8.6$ $\upmu$m in (d,e,f). Normalised logarithmic colour scales are used with the corresponding colour bar on the bottom of each image. The black lines on the bottom right of (a,d) represent 15 $\upmu$m and in (b,e) represent 1 $\upmu$m$^{-1}$ scale bars. The red vertical dashed lines in (c \& f) symbolise the boundaries of the reduced Brillouin zone of the polariton crystal.}
	\label{Fig_BlochBand_ExpChain}
\end{figure}

\section*{Results}
To demonstrate the fundamental applicability of this technique for band-structure engineering we start with a periodic ring-shaped arrangement of narrow Gaussian pumps (full-width-half-maximum $\approx 2$ $\upmu$m) operating at 50\% above the excitation density required for the formation of a macroscopic coherent single-particle state (see Figure~\ref{Fig_BlochBand_ExpOctagon}). The ring-shaped polariton condensate is characterised by bright emission spots at the location of the pumps and standing wave fringes in-between. In Figs.~\ref{Fig_BlochBand_ExpOctagon}(a,b) and (c,d) we show the experimental real-space and $k$-space photoluminescence (PL) distributions for a regular and dimerised octagon pattern of pump spots respectively. In cold-atom systems, periodic lattice potentials are typically superposed with harmonic trapping potentials, affecting the resulting finite band structure. Here, we engineer ring-shaped exciton-polariton crystals that more closely resemble ideal periodic bands of solid-state crystals, avoiding also any undesired edge effects appearing in finite linear condensate chains.

Another lattice geometry of fundamental interest is that of a linear chain where in Fig.~\ref{Fig_BlochBand_ExpChain} we show the experimental real-space and $k$-space PL distributions along with the corresponding dispersions (energy resolved $k$-space along the chain) for eight pumps with lattice constants of (a, b \& c) $a \approx 13$ $\upmu$m and (d, e \& f) $a \approx 8.6$ $\upmu$m. One can see in Fig.~\ref{Fig_BlochBand_ExpChain}(c,f) that the condensate chain dominantly occupies the high-symmetry points of the reduced Brillouin zone and all repeated zones within the free polariton dispersion. These results evidence that polaritons generated at the pump spots sense the periodic nature of the potential, resulting in macroscopic coherent Bloch states even for relatively few pump cells, thus qualifying the technique. The energy band wherein the system condenses into can be controlled by changing the separation between neighbouring condensates as is demonstrated in Fig.~\ref{Fig_BlochBand_ExpChain}, where we realise access to non-linear condensate dynamics in arbitrarily excited states through all-optical control.

We note the intricate Talbot interference patterns observed experimentally in the regions perpendicularly away from the chains, e.g., in Fig.~\ref{Fig_BlochBand_ExpChain}a. Such patterns were previously demonstrated for polariton condensates using a chain of etched mesa traps~\cite{gao_talbot_2016} and demonstrates the ability of optically imprinted condensates with the concomitant potentials to achieve effects of etched/patterned systems. Moreover, polaritons condensing into the high-symmetry points of the lattice, observed also in Refs.~[\onlinecite{lai_coherent_2007, kim_dynamical_2011, Winkler_PRB2016}], can be intuitively understood from the fact that these Bloch modes have the strongest overlap with the gain (pump) region. The results are verified both through diagonalisation of the non-Hermitian Bloch problem, and by numerically solving the driven-dissipative Gross-Pitaevskii equation describing a coherent macroscopic field of polaritons under pumping and dissipation [see Sections S1 and S2 in the Supplementary Information (SI)].
\begin{figure*}
	\centering
	\includegraphics[width=17.2cm]{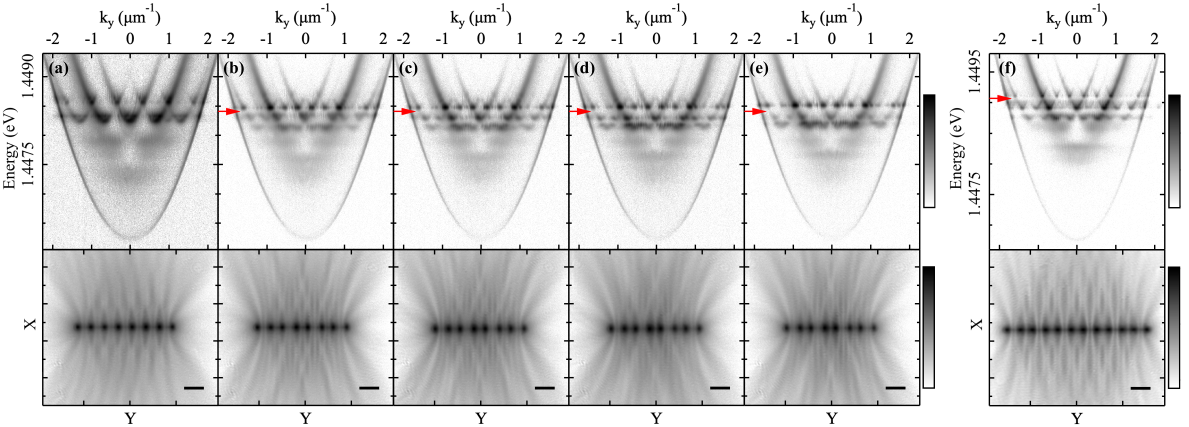}
	\caption{Experimental dispersion (top) and real-space intensity distribution (bottom) of the PL from chains of eight polariton condensates with alternating separation distances where $a_{l}\approx10.4$ $\upmu\mathrm{m}$ and (a) $a_{s}\approx10.4$ $\upmu\mathrm{m}$, (b) $a_{s}\approx9.0$ $\upmu\mathrm{m}$, (c) $a_{s}\approx8.9$ $\upmu\mathrm{m}$, (d) $a_{s}\approx8.7$ $\upmu\mathrm{m}$ and (e) $a_{s}\approx8.5$ $\upmu\mathrm{m}$. Panel (f) shows the PL dispersion (top) and real-space intensity distribution (bottom) for a chain of twelve condensates with $a_{l}\approx10.2$ $\upmu\mathrm{m}$ and $a_{s}\approx9.2$ $\upmu\mathrm{m}$. The horizontal bars in the bottom right corner of each real-space distribution correspond to $15$ $\upmu\mathrm{m}$.}
	\label{Fig_BlochBandSplitting}
\end{figure*}

In cold-atom systems topologically non-trivial band structure can be engineered by generating artificial gauge potentials using laser beams, where the hopping amplitude between adjacent lattice sites picks up a controllable phase factor (Peierls substitution) from the laser amplitudes~\cite{Ruostekoski_PRL2002, Jaksch_NJP2003, Jimenez_PRL2012} or from periodic modulation~\cite{Kolovsky_EuLett2011, Jotzu_Nature2014}. Here we engineer an alternating pattern of tunnelling amplitudes between neighbouring condensates by utilising the variation of the hopping amplitude with the condensate separation, such that the interference between the neighbouring sites is staggered. Our approach provides increased flexibility for engineering topologically non-trivial systems, with both periodic boundary conditions for a ring-shaped lattice (see Fig.~\ref{Fig_BlochBand_ExpOctagon}) and open boundary conditions for a linear chain, similar to the SSH model~\cite{Su_PRL1979} of polyacetylene polymer chains where alternating hopping amplitudes $J_\pm$ split the energy bands and generate edge states.

Figure~\ref{Fig_BlochBandSplitting} shows the experimental dispersion (top) and real-space PL distribution (bottom) for chains of eight condensates, demonstrating the splitting and periodic doubling of the band as the difference between the long ($a_{l}$) and short separation ($a_{s}$) is increased (panels $\mathrm{a}\rightarrow\mathrm{e}$). For marginal differences in separation distance, $\Delta = a_{l} - a_{s}$, the band gap formed is smaller than or comparable to the linewidths of the condensate polaritons and thus not fully resolvable. Increasing $\Delta$  leads to an increased band splitting and the gaps become clearly visible when they exceed the polariton linewidth. Eventually the band splitting becomes significant enough that adjacent energy bands mix; see Fig.~\ref{Fig_BlochBandSplitting}. Figure~\ref{Fig_BlochBandSplitting}(f) shows the dispersion and real-space of the PL from a chain of twelve condensates with $a_{l}=10.2$ $\upmu\mathrm{m}$ and $a_{s}=9.2$ $\upmu\mathrm{m}$. The increased number of unit cells results in the band features becoming sharper and thus the extinction within the newly opened band gap is increased.
\begin{figure}[t!]
	\centering
	\includegraphics[width=12cm]{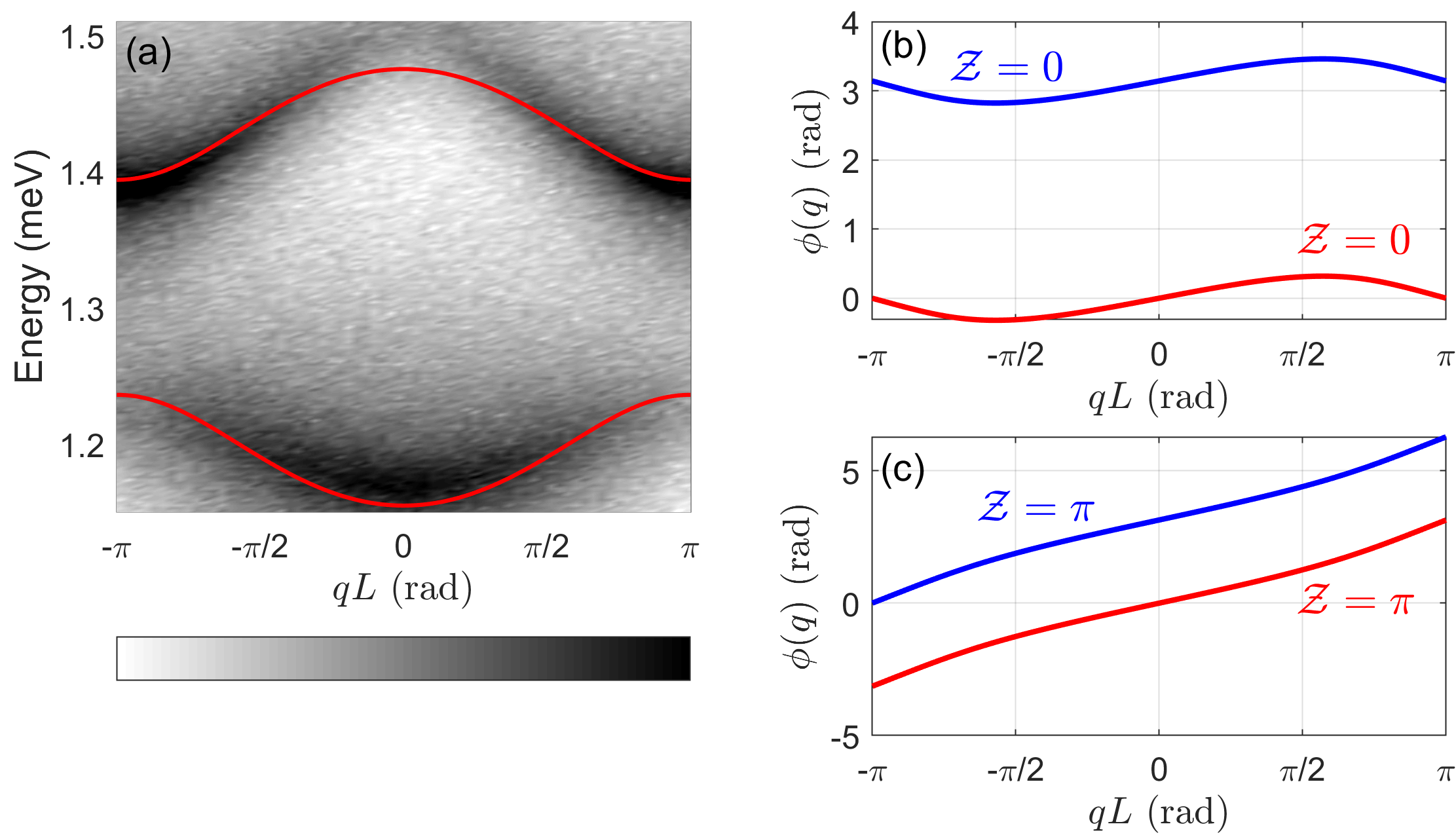}
	\caption{(a) Grayscale colourmap shows the numerically time resolved dispersion in a complex Gaussian potential lattice representing the experiment in Fig.~\ref{Fig_BlochBandSplitting}f. The broadening of the bands is a consequence of including the polariton linewidth in the calculation, estimated as $\hbar\gamma=0.12$ meV (see Section S1 in SI). Zero energy represents bottom of the lower polariton dispersion. Red curves are calculated energies from Eq.~\eqref{eq.schroMain} for a periodic lattice in the basis of crystal momentum $q$ [see Eq.~(S14) in SI]. (b,c) Show the polariton Bloch state phase $\phi(q)$ in the conduction (red) and valence (blue) band corresponding to $\delta = \pm0.5$ $\upmu$m respectively. Other parameters are:  $d = 9.5$ $\upmu$m, $k_c = 1.5$ $\upmu$m$^{-1}$, $\mu = 0.32$ meV ps$^2$ $\upmu$m$^{-2}$, $\eta = 0.24$, $\Omega = 1.315$ meV, and $V_0 = 1.44 + i0.5$ meV. Here $L=2d$ denotes the unit cell size and $\mathcal{Z}$ the calculated Zak phase [see Eq.~(S16) in SI].}
	\label{fig.top}
\end{figure}

The strong localisation of the polaritons at their respective excitation spots permits description by a discretised tight-binding picture. As distances between adjacent condensates are staggered their coupling strength follows suit due to both differences in travel times (i.e., the condensate envelope decays rapidly outward from its respective pump spot) and interference coming from their large $k$-vector. Projecting the system to a basis of gain-localised orbitals (see Section S3 in SI) and neglecting polariton-polariton interactions, one obtains for weak coupling between adjacent sites,
\begin{equation} \label{eq.schroMain}
i\hbar \dot{c}_n  =  \Omega c_n  +  \sum_{\langle nm \rangle} J(k_c|x_n-x_m|) c_{m}.
\end{equation}
Here, $\Omega$ is the complex-valued potential energy of polaritons generated at their respective pump spots, $J$ is a function describing the coupling, $k_c$ is the outflow momentum of the polaritons from their pump spot, and $|x_n-x_m|$ is the distance between pump spots. For the dimerised system, the sum is taken over nearest neighbours and one obtains a single-particle two-band problem quantified in terms of two hopping amplitudes,
\begin{align} \notag \label{eq.JMain}
J_\pm = J(k_c[d\pm\delta])= & \eta \left(V_0\cos{(k_c [d\pm\delta])} - \frac{\hbar^2k_c}{\mu} \sin{(k_c [d\pm\delta])} \right) \\
& \times |H_0^{(1)}(k_c[d\pm\delta])|.
\end{align}
Here, $d\pm\delta$ represent the long and short distances between the pumps respectively, and $H_0^{(1)}$ is the zeroth order Hankel function of the first kind, $\mu$ is the polariton mass, $V_0$ the strength of the complex valued pump induced potential, and $\eta$ a fitting parameter. The physical meaning of Eq.~\eqref{eq.schroMain} is that polaritons do not tunnel from one site to the next (evanescent coupling) but rather ballistically exchange energy with polaritons at the neighbouring pump spots. The term \emph{ballistically} refers to the presence of the sine and cosine functions which originate from the interference of gain localised wavefunctions with outflow momentum $k_c$. Equation~\eqref{eq.JMain} represents a non-Hermitian version of the SSH model~\cite{Su_PRL1979}. As the system possesses inversion symmetry in two distinct choices of origin, that is, one of either sublattice, its geometric phase (better known as Zak's phase for one dimensional systems~\cite{Zak_PRL2989}) can only take on values $0$ or $\pi$ (modulo $2\pi$) when the origin is chosen at an inversion center. Moreover, topologically protected defect (gap) states will appear by breaking the translational symmetry of the lattice, either at the finite system boundaries or by introducing a defect in a periodic structure as depicted in Fig.~\ref{Fig_BlochBand_ExpOctagon} (see Section S2.3 in SI).

In Fig.~\ref{fig.top} we present numerical results reproducing the experimental gap opening shown in Fig.~\ref{Fig_BlochBandSplitting}f. Figure~\ref{fig.top}a shows the fitted gapped bulk dispersion from Eq.~\eqref{eq.schroMain} (red curves) in the lattice Brillouin zone with $d = 9.5$ $\upmu$m, $\delta = 0.5$ $\upmu$m, and $k_c = 1.5$ $\upmu$m$^{-1}$. The curves are plotted on top of a black-and-white colourmap showing the numerically time-resolved single-particle dispersion based on a Monte Carlo technique (see Section S1 in SI). We point out that even for a dimerised lattice ($\delta \neq 0$) the gap can be closed by only tuning $k_c$ which changes the interference between condensates. For the parameters given in the caption of Fig.~\ref{fig.top} it closes when, e.g., $k_c \approx 1.395$ $\upmu$m$^{-1}$. In Fig.~\ref{fig.top}b,c we plot the phases of the polariton Bloch states across the Brillouin zone in the conduction (red) and valence (blue) band (see Section S2.3, SI). Figure~\ref{fig.top}b,c corresponds to the two different dimerisations $\delta = \pm0.5$ $\upmu$m respectively. Integrating $\phi(q)$ across the Brillouin zone shows a $\pi$ change in the Zak phase between the dimerisations, marking the existence of two topologically distinct phases. The findings are corroborated through first principle calculations on the polariton system Schr\"{o}dinger equation (see Section S2.2, SI). We point out that our system is very different from that of hybridised orbitals in micropillar chains~\cite{StJean_NatPho2017}, where in the current case, the opening of the gap arises from the staggered interference between adjacent polariton condensate ``antennas''.
\begin{figure}
	\centering
	\includegraphics[width=8.6cm]{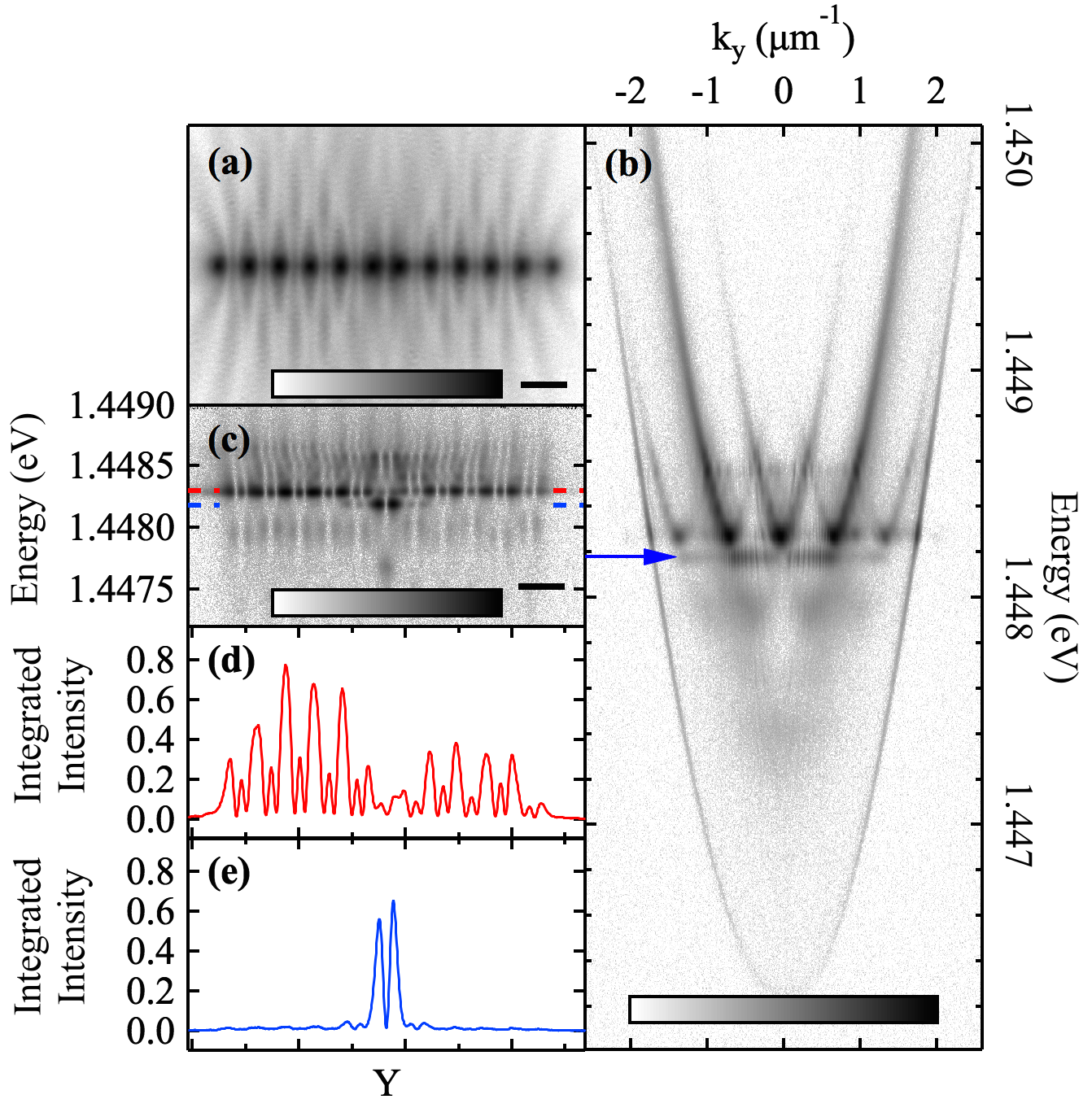}
	\caption{Chain of twelve polariton condensates with $a\approx10.2$ $\upmu\mathrm{m}$ and $a_{d}\approx9.0$ $\upmu\mathrm{m}$. (a) Experimental real-space PL intensity distribution, (b) dispersion, (c) energy resolved strip of real-space, (d \& e) line profiles across energy resolved real-space centred around the red and blue dashed lines in (c) respectively. The horizontal solid black lines in (a \& c) correspond to $15\upmu\mathrm{m}$ and the normalised logarithmic colour scales are shown at the bottom of each colour-map.}
	\label{Fig_BlochBandDefectState}
\end{figure}

If the translational symmetry is broken then gap (defect) states appear. These manifest as flat bands within the dispersion, showing strong spatial localisation around the position of the defect in the pump geometry at the defect state energy. Figure~\ref{Fig_BlochBandDefectState}a shows the experimental real-space PL distribution from a chain of 12 condensates with separation distances of $a\approx10$ $\upmu\mathrm{m}$ except between the central two pump spots where the separation is reduced to $a_{d}\approx8.8$ $\upmu\mathrm{m}$, creating a defect in the potentials periodicity. A corresponding flat band is visible in the dispersion (indicated with the blue arrow in Fig.~\ref{Fig_BlochBandDefectState}b) and the energy resolved strip of real-space (Fig.~\ref{Fig_BlochBandDefectState}c) demonstrates strong localisation of the condensate occupying the defect energy (Fig.~\ref{Fig_BlochBandDefectState}e). Such states are the optically generated analog of polariton gap solitons observed in~\cite{Tanese_NatComm2013}. The rest of the condensate which occupies the higher energy band suffers significant suppression around the defect (see Fig.~\ref{Fig_BlochBandDefectState}d). This suppression is a consequence of the bulk energy bands vanishing around the defect and thus inhibiting energy flow between the left and the right bulk region of the optical polariton crystal. We present simulations on such defect states in Section S4 in the SI.
\begin{figure*}
	\centering
	\includegraphics[width=17.2cm]{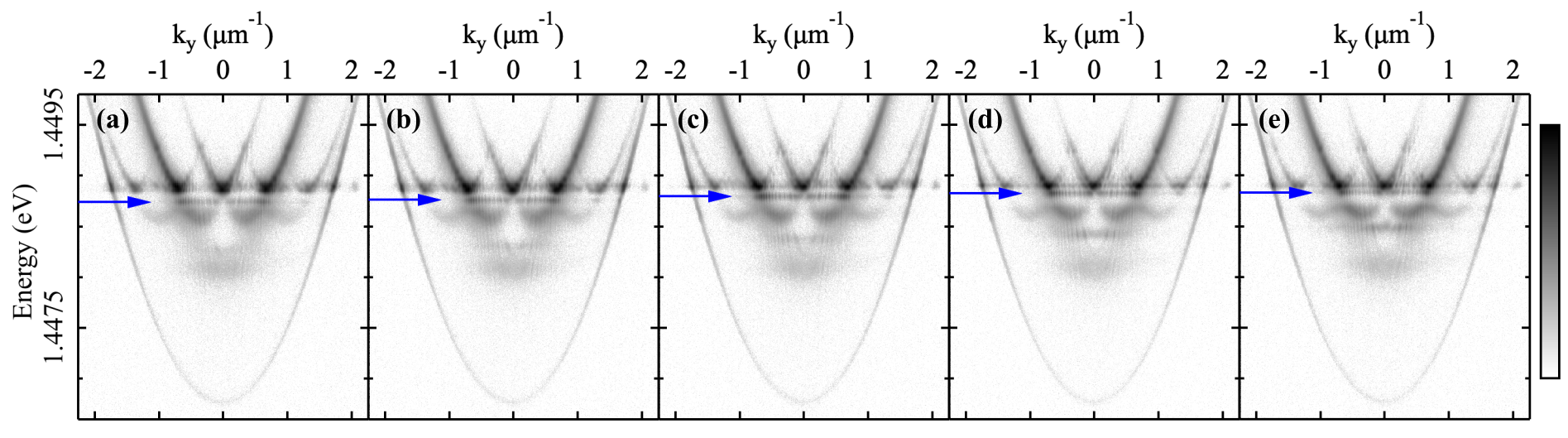}
	\caption{Experimental dispersions of the PL from chains of twelve polariton condensates with $a\approx10$ $\upmu\mathrm{m}$ and a defect length of (a) $a_{d}\approx9.0$ $\upmu\mathrm{m}$, (b) $a_{d}\approx8.6$ $\upmu\mathrm{m}$, (c) $a_{d}\approx8.2$ $\upmu\mathrm{m}$, (d) $a_{d}\approx7.6$ $\upmu\mathrm{m}$ and (e) $a_{d}\approx7.1$ $\upmu\mathrm{m}$. Each colour-map uses the normalised logarithmic colour scale shown to the right of (e).}
	\label{Fig_BlochBandDefectSpectralControl}
\end{figure*}

Optically imprinting the potential landscape affords the ability to finely tune the spectral position of the defect state, within the gap, by only changing the defect length ($a_{d}$) in the excitation geometry. The PL dispersions for chains of twelve condensates with $a=10$ $\upmu\mathrm{m}$ for five defect lengths between $a_{d}=8.9$ $\upmu\mathrm{m}$ and $a_{d}=7.1$ $\upmu\mathrm{m}$ are shown in Fig.~\ref{Fig_BlochBandDefectSpectralControl}(a-e). As the defect separation distance is reduced the resulting flat band blueshifts from towards the bottom of the gap to the top, at which point it begins to mix with neighbouring energy bands. We note that there also exists a flat band in the lower energy band gap that also demonstrates the same blueshift behaviour with reducing defect length.

\section*{Discussion}

Our study advances the emulation of many different lattice structures using a recyclable, and optically reprogrammable, multi-purpose platform in the strong light-matter coupling regime. The controllable condensation into arbitrarily excited Bloch states of the system, gives access to excited orbital many-particle dynamics which previously have been difficult to reach in solid-state systems. In particular, we overcome the challenge of realising a condensate lattice with periodic boundary conditions which, in general, is attractive for analytical considerations (Bose-Hubbard model on a ring), and more closely resembles classic band-structure models of solid-state physics. The observed defect-state condensation paves the way towards strong nonlinear lattice physics, with application in polaritonic devices such as information routing, and fine tunable emission wavelength lasers. Additionally, we expect that topological defect lasing can be realised by adjusting the lattice geometry. We point out that the current study is performed in the scalar polariton regime, but can be easily extended to include its spin degree of freedom by changing the polarization of the pump. By extending to spin lattice physics, one can in principle design optical Chern insulators given the inherent spin-orbit coupling of polaritons in conjunction with applied magnetic fields. An exciting area then for future research is expanding to topologically protected transport states with investigation into robustness against engineered imperfections.

\section*{Methods}
We use a planar distributed Bragg reflector microcavity with a $2\lambda$ GaAs based cavity containing eight $6$ nm InGaAs QWs organised in pairs at the three anti-nodal positions of the confined field, with an additional QW at the final node either side of the cavity \cite{cilibrizzi_polariton_2014}. The sample is cooled to $\sim6$ K using a cold finger flow cryostat and is excited with a monomode continuous wave laser blue detuned energetically above the stop band to maximise coupling in efficiency. The laser is modulated in time into square wave packets with a frequency of $10$ kHz and a duty cycle $<5\%$ to prevent sample heating. The spatial profile of the excitation beam is sculpted using a phase-only spatial light modulator to imprint a phase map so that, when the beam is focused via a 0.4 numerical aperture microscope objective lens, the desired real-space is projected onto the samples surface. The same objective lens is used to collect the photoluminescence which is then directed into the detection setup.

By controlling the spatial intensity distribution of the non-resonant excitation beam we imprint a reprogrammable potential landscape~\cite{tosi_sculpting_2012,tosi_geometrically_2012,berloff_realizing_2017} without the need of irreversible engineering. In the relaxation process from a non-resonant optical injection of free charge carriers to the polariton condensate, an incoherent `hot' excitonic reservoir is produced that feeds the condensate. This reservoir is co-localised with the non-resonant excitation beam(s) and due to the strong polariton-exciton interaction results in a potential hill for polaritons where the excitation density is high~\cite{Wertz_NatPhys2010}. This method additionally enables the elimination of large inhomogeneities since each element of the potential lattice can be adjusted through the power or shape of its respective pump element, such that the system achieves a homogeneous crystal structure.

\section*{Acknowledgements}
L.P, H.S. and P.G.L. acknowledge the support of the UK’s Engineering and Physical Sciences Research Council (grant EP/M025330/1 on Hybrid Polaritonics).

\section*{Author contributions}
P.G.L. led the research project. P.G.L. and L.P designed the experiment. L.P. carried out the experiments and analysed the data. H.S. and J.R. developed the theoretical modelling. H.S performed numerical simulations. All authors contributed to the writing of the manuscript.

\section*{Competing interests}
The authors declare no competing interests.


\newpage

\setcounter{equation}{0}
\setcounter{figure}{0}
\setcounter{table}{0}
\setcounter{page}{1}
\renewcommand{\theequation}{S\arabic{equation}}
\renewcommand{\thefigure}{S\arabic{figure}}
\renewcommand{\thesection}{S\arabic{section}}

\begin{center}\large\textbf{Supplementary Information}\end{center}

\section{Driven-dissipative Gross-Pitaevskii theory} \label{sec1}
The transition from a thermally stochastic state to a macroscopic coherent polariton condensate can be captured within the mean field theory approach. The condensate order parameter $\Psi(\mathbf{r},t)$ is then described by a two-dimensional (2D) semiclassical wave equation referred as the {\it driven-dissipative Gross-Pitaevskii equation} (dGPE) coupled with an excitonic reservoir which feeds non-condensed particles to the condensate~\cite{Wouters_PRL2007}. The reservoir is divided into two parts: An {\it active} reservoir $n_A(\mathbf{r},t)$ belonging to excitons which experience bosonic stimulated scattering into the condensate, and an {\it inactive} reservoir $n_I(\mathbf{r},t)$ which sustains the active reservoir~\cite{Lagoudakis2010a, Lagoudakis_PRL2011}.
\begin{subequations} \label{eq.GP_Res}
	\begin{align} \label{eq.GP}
	i  \frac{d \Psi}{d t} & = \left[ -\frac{\hbar \nabla^2}{2m} + g(n_A+n_I) + \alpha |\Psi|^2 + \frac{i }{2} \left( R n_A - \gamma \right) \right] \Psi,\\ \label{eq.ResA}
	\frac{ dn_A}{d t} & = - \left( \Gamma_A + R |\Psi|^2 \right) n_A + W n_I, \\ \label{eq.ResI}
	\frac{ d n_I}{d t} & = -  \left(\Gamma_I + W \right) n_I + P(\mathbf{r}).
	\end{align}
\end{subequations} 
Here, $m$ is the effective mass of a polariton in the lower dispersion branch, $\alpha$ is the interaction strength of two polaritons in the condensate, $g$ is the polariton-reservoir interaction strength, $R$ is the rate of stimulated scattering of polaritons into the condensate from the active reservoir, $\gamma$ is the polariton condensate decay rate, $\Gamma_{A,I}$ are the decay rates of active and inactive reservoir excitons respectively, $W$ is the conversion rate between inactive and active reservoir excitons, and $P(\mathbf{r})$ is the non-resonant continuous wave pump profile. The polariton mass and lifetime are based on the sample properties: $m = 0.32$ meV ps$^2$ $\upmu$m$^{-2}$ and $\gamma = 1/5.5$ ps$^{-1}$. We choose values of interaction strengths typical of GaAs based systems: $\hbar \alpha = 1.6$ $\upmu$eV  $\upmu$m$^2$ and $g = 2 \alpha$. The non-radiative recombination rate of inactive reservoir excitons is taken here much smaller than the condensate decay rate $\Gamma_I^{-1} = 500$ ps whereas the active reservoir is taken comparable to the condensate lifetime $\Gamma_A^{-1} = 5$ ps due to fast thermalisation to the exciton background~\cite{Wouters_2008PRB} and partly because active excitons are within the light cone. The final two parameters are then found by fitting numerical results to experiment. This gives the values $\hbar R = 36.3$ $\upmu$eV $\upmu$m$^{-2}$, and $W = 0.05$ ps$^{-1}$.

The pump is written,
\begin{equation} \label{eq.pump}
P(\mathbf{r}) = P_0 f(\mathbf{r}) = P_0 \sum_{n=1}^N \exp{[-(x^2 + y_n^2)/2w^2]}.
\end{equation}
where $N$ is the number of pumps in question and $y_n$ their coordinates along the $y$-axis, $P_0$ denotes the pump power density and $w$ corresponds to a 2 $\upmu$m full-width-half-maximum (FWHM).

\subsection{Time resolved dispersion} \label{sec.stoch_disp}
Here we discuss a method to numerically time-resolve the dispersion of the Schr\"{o}dinger equation in some arbitrary potential $V(\mathbf{r})$. The linear parts of the dGPE form a non-Hermitian Schr\"{o}dinger equation written,
\begin{equation} \label{eq.schro}
i  \frac{d \Psi}{d t}  = \left[ -\frac{\hbar \nabla^2}{2m} + g(n_A+n_I)  + \frac{i }{2} \left( R n_A - \gamma \right) \right] \Psi.
\end{equation}
For nonzero pump powers $P_0\neq0$, and in the linear regime ($|\Psi|^2 \simeq 0$), the reservoirs converge to a steady state solution satisfying $\dot{n}_{A,I} = 0$. We can then write an effective potential,
\begin{equation}
V(\mathbf{r})/\hbar = g(n_A+n_I)  + \frac{i R n_A}{2} = \left[g \left(1 + \frac{W}{\Gamma_A}\right) + \frac{iR}{2}\right] \frac{P(\mathbf{r})}{\Gamma_I + W}.
\end{equation}
For brevity we will write $V(\mathbf{r}) = V_0 f(\mathbf{r})$ where $V_0$ is defined,
\begin{equation}
V_0 = \hbar \left[g \left(1 + \frac{W}{\Gamma_A}\right) + \frac{iR}{2}\right] \frac{P_0}{\Gamma_I + W}.
\end{equation}
Equation~\eqref{eq.schro} then becomes
\begin{equation} \label{eq.schro}
i  \hbar \frac{d \Psi}{d t}  = \left[ -\frac{\hbar^2 \nabla^2}{2m} + V(\mathbf{r}) - \frac{i\hbar \gamma}{2} \right] \Psi.
\end{equation}
The dispersion of Eq.~\eqref{eq.schro} can be extracted by two means. Firstly, by introducing a white noise term $d\theta$ based on a quantum Monte Carlo technique in the Wigner representation~\cite{walls_quantum_2008},
\begin{equation} \label{eq.schro_W}
i  \hbar \frac{d \Psi}{d t}  = \left[ -\frac{\hbar^2 \nabla^2}{2m} + V(\mathbf{r}) - \frac{i\hbar \gamma}{2} \right] \Psi + \frac{d \theta}{dt},
\end{equation}
where the white noise correlation functions are written $\langle d\theta(x_i) d\theta(x_j) \rangle  =0 $ and $\langle d\theta^*(x_i) d\theta(x_j) \rangle  =2 \delta(x_i - x_j) dt$. 
The dynamics of many random fields generated by $d\theta$ can then be averaged over in $k$-space ($k_x$-$k_y$ plane) to produce the dispersion experienced by particles in the system. This method is particularly useful for potentials which lack the usual symmetries required by Bloch theorem to work. Alternatively, if the system possesses translational symmetry then Bloch's theorem can be applied (see Section~\ref{sec.bloch}).
\begin{figure}[t!]
	\centering
	\includegraphics[width=1\columnwidth]{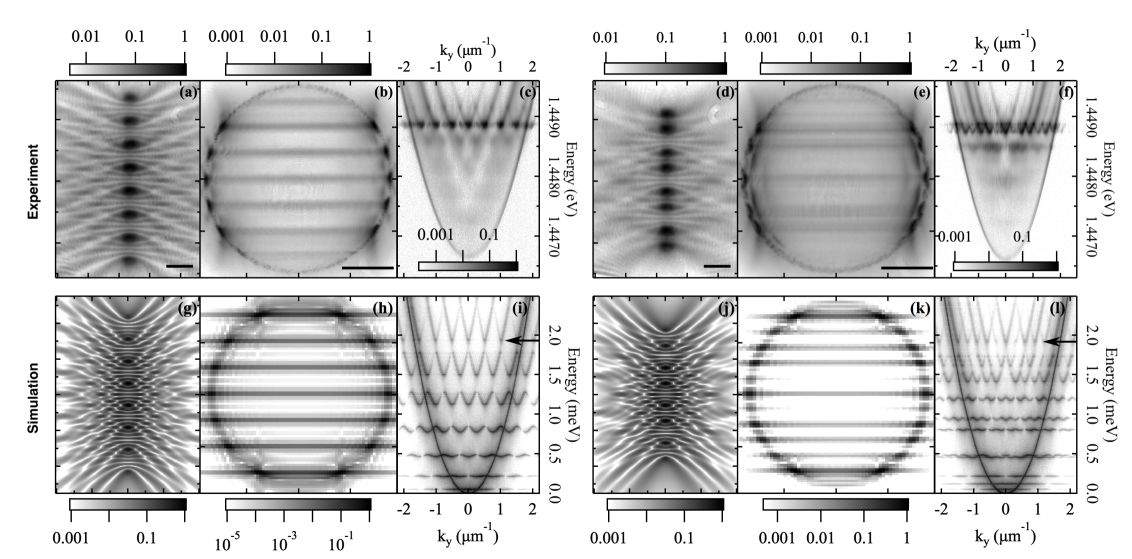}
	\caption{(a,d) Experimental PL from a chain of eight polariton condensates in real space, (b,e) $k$-space, and (c,f) the dispersion. The lattice constant is $a=13$ $\upmu$m in (a,b,c) and alternating with $a_s = 8.6$ $\upmu$m and $a_l = 13$ $\upmu$m in (d,e,f). The solid lines on the bottom right of (a,d) represents 15 $\upmu$m and in (b,e) represents 1 $\upmu$m$^{-1}$. In the lower row we show the corresponding steady state condensate wavefunction obtained from simulating Eq.~\eqref{eq.GP_Res} in (g,j) real space and (h,k) $k$-space. Panels (c,f) show the time resolved single-particle dispersion of the system for $V_0 = 1.44$ meV and $\gamma = 0$. The energy of the simulated condensates in (g,h) and (j,k) is found to be $E \sim 1.9$ meV which lies in the high-symmetry points of the 7th and 11th bands in their respective dispersions (marked with black arrows). Pump power is set to $P_0 = 19$ ps$^{-1}$ $\upmu$m$^{-2}$.}
	\label{sim1}
\end{figure}

\subsection{dGPE results}
Comparison between experiment and simulation is shown in Fig.~\ref{sim1}. The upper row shows experiment and lower row the simulation. Figure~\ref{sim1}(a,b,c) shows the condensate real-space, $k$-space, and Fourier space photoluminescence (PL) for a chain of $N = 8$ pump spots with uniform spacing between them, $a_l = a_s = 12.4$ $\upmu$m. Figure~\ref{sim1}(d,e,f) shows the same for a chain of $N=8$ pumps with two interchanging separation distances between the pump spots,  $a_l = 12.4$ $\upmu$m and $a_s = 8.8$ $\upmu$m. In Fig.~\ref{sim1}(g,h) we plot the simulated steady state of the condensate order parameter showing the real-space density $|\Psi(\mathbf{r})|^2$ and $k$-space density $|\hat{\Psi}(\mathbf{k})|^2$ respectively for the uniformly $N=8$ spaced pump spots, $a_l = a_s = 12.4$ $\upmu$m. In Fig.~\ref{sim1}(j,k) we plot the simulated steady state of the condensate order parameter showing the real-space density $|\Psi(\mathbf{r})|^2$ and $k$-space density $|\hat{\Psi}(\mathbf{k})|^2$ respectively for the $N=8$ pumps with two interchanging separation distances, $a_l = 12.4$ $\upmu$m and $a_s = 8.8$ $\upmu$m. Figures~\ref{sim1}(i,l) show the numerically time-resolved single-particle dispersion (see Section~\ref{sec.stoch_disp}) for the corresponding $N=8$ uniform and dimerised chains of Gaussian potentials, $V(\mathbf{r}) = V_0 f(\mathbf{r})$, as given by Eq.~\eqref{eq.pump}. Both dispersions are numerically obtained for $V_0 = 1.44$ meV and $\gamma = 0$ in order to avoid band-broadening and to show the low energy band structure more clearly. The energy of the simulated condensates in Fig.~\ref{sim1}(g,h) and (j,k) is found to be $E \sim 1.9$ meV which lies in the high-symmetry points of the 7th and 11th band in their respective dispersions (marked with black arrows).
\begin{figure}[t!]
	\centering
	\includegraphics[width=0.8\columnwidth]{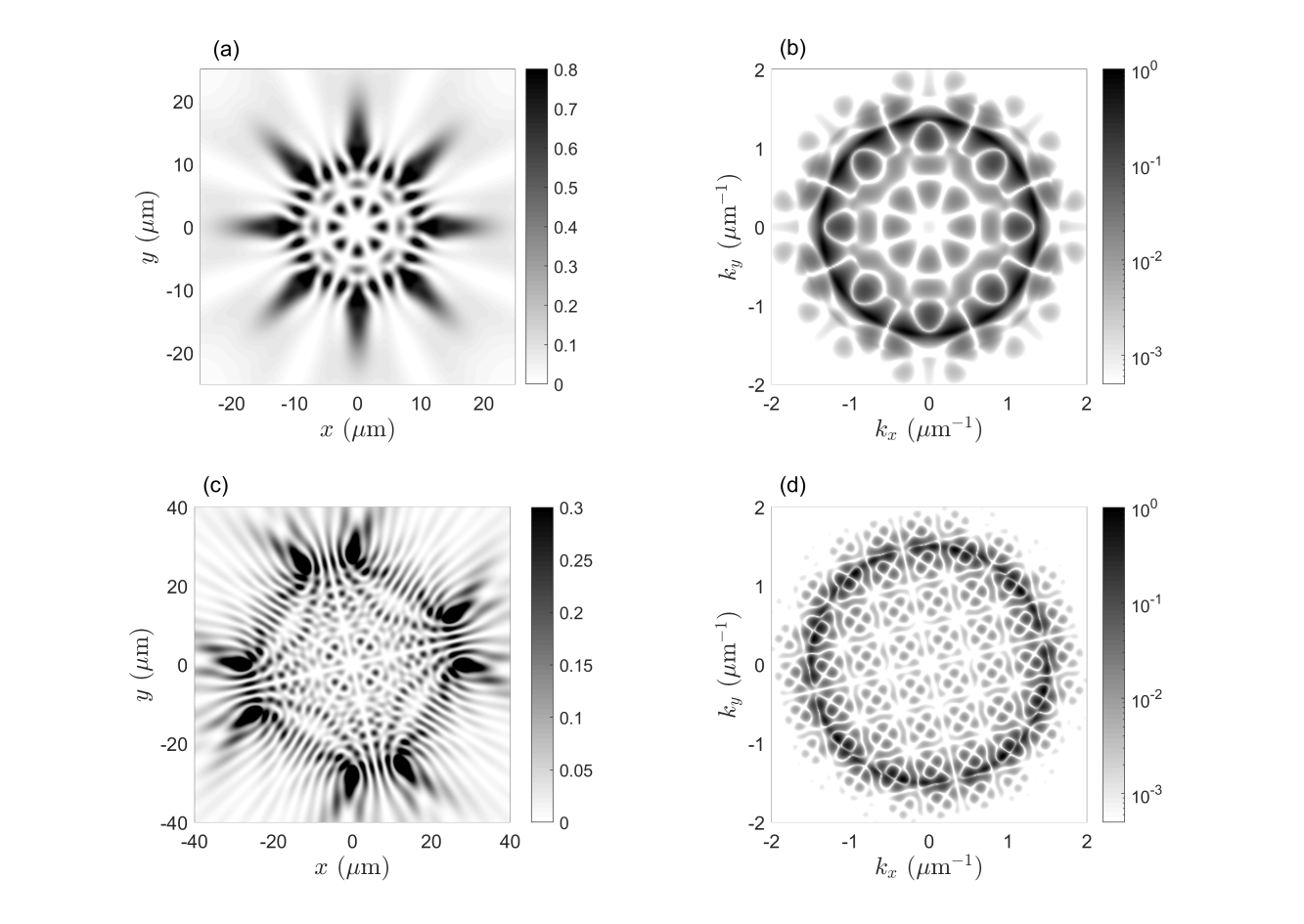}
	\caption{Simulated steady state condensate (a,c) real-space density $|\Psi(\mathbf{r})|^2$ and (b,d) $k$-space density $|\hat{\Psi}(\mathbf{k})|^2$ for a regular and dimerised octagon pump pattern respectively. Densities in all panels are normalized to unity. Panels (a,c) are linear in colorscale and saturated at 80\% and 30\% respectively to show more clearly the underlying interference pattern. Panels (b,d) are shown in a logarithmic scale.}
	\label{fig.octagon}
\end{figure}

In Fig.~\ref{fig.octagon}(a,c) we show the real-space and (c,d) $k$-space densities of the simulated steady state condensate order parameter $\Psi(\mathbf{r})$ for a regular ($P_0 = 15$ ps$^{-1}$ $\upmu$m$^{-2}$) and dimerised ($P_0 = 19$ ps$^{-1}$ $\upmu$m$^{-2}$) octagon pump pattern respectively. The simulations illustrate the good agreement between the dGPE and the observed experimental PL (see Fig.~1 in main text).

\section{Continuum Bloch theorem} \label{sec.bloch}
Polariton condensation into the high-symmetry points of the Brillouin zone can be understood by scrutinising the linear part of the dGPE [see Eq.~\eqref{eq.schro}] in the pump lattice bulk by assuming that the system is infinite and periodic such that $V(y)  =V(y+L)$ with $L$ being the lattice constant. We can then apply Bloch's theorem where we write the polariton single particle wavefunction in the factorised form of crystal momentum $k_y = q$ along the pump chain and Bloch states in the $n$th band $u_{n,q}(y)$,
\begin{equation} \label{eq.bloch}
\Psi_n(y) = e^{i q y} u_{n,q}(y).
\end{equation}
Here, $u_{n,q}(y) = u_{n,q}(y+L)$, and the $x$-coordinate notation is dropped since the system is only periodic along $y$. Substituting Eq.~\eqref{eq.bloch} into Eq.~\eqref{eq.schro} the Bloch eigenvalue problem then reads,
\begin{equation}\label{eq.bloch_eig}
\left[ \frac{\hbar^2}{2m}\left(q - i \frac{\partial}{\partial y}\right)^2 + V(y) -  \frac{i\hbar \gamma}{2} \right] u_{n,q}(y) = E_{n,q} u_{n,q}(y).
\end{equation}
We can diagonalise Eq.~\eqref{eq.bloch_eig} by expressing the Gaussian potential as a Fourier series in the basis of the reciprocal lattice vectors $G =2\pi/L$~\cite{Kittel_2004},
\begin{equation}
V(y) = \sum_{G} V_{G} e^{iG y}.
\end{equation}
\begin{figure}[t!]
	\centering
	\includegraphics[width=0.5\columnwidth]{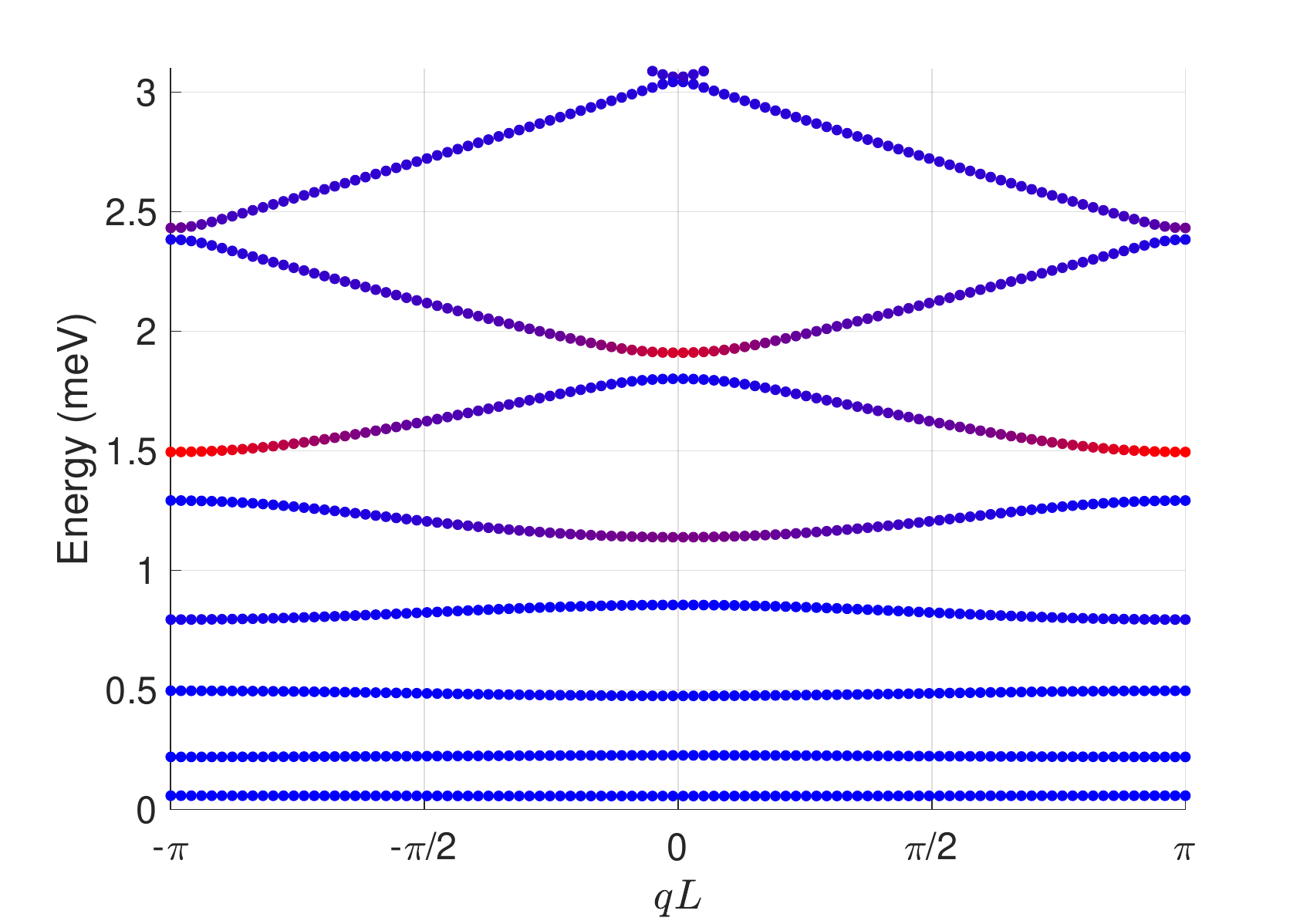}
	\caption{Calculated bands using Bloch's theorem for complex valued Gaussian potentials in a uniform lattice with parameters: $L = 12.4$ $\upmu$m, $w = 0.85$ $\upmu$m, $V_0 = 1.44 + i 0.5$, $m = 0.32$ meV ps$^2$ $\upmu$m$^{-2}$ and $\gamma = 1/5.5$ ps$^{-1}$. The colorscale indicates crystal momenta with low (blue) and high (red) optical gain, i.e. positive imaginary energy.}
	\label{fig_bloch}
\end{figure}

\subsection{Condensation in high-symmetry points}  \label{sec.bloch_cond}
The top eight bands $E_{n,q}$ in the Brillouin zone of the system for 
$V_0 = 1.44+i0.5$ meV are shown in Fig.~\ref{fig_bloch}. The value of $V_0$ is extracted from gDPE simulations previously shown in Fig.~\ref{sim1}. The colorscale denotes energies with low imaginary part (blue) and large imaginary part (red). We stress that the imaginary part of the energies $E_{n,q}$ corresponds directly to the gain of the Bloch waves. Energies of highest gain are found in the $\Gamma$ and $M$ points of the Brillouin zone of positive effective mass, which can be intuitively understood from the Bloch states in these points having maximum overlap with the gain (pump) region. The fact that bands 6 and 7 possess optimal gain can be understood by inferring that lower energy Bloch waves have poor penetration (overlap) into the repulsive potentials which contain all the gain, whereas higher energy Bloch waves transmit too easily through the gain regions. Therefore, only certain bands will have ``ideal'' penetration into the gain region of the pump induced potentials. As a consequence, condensation into the crystal bands high-symmetry points can be predicted by straightforward diagonalisation of the Schr\"{o}dinger equation. A similar observation was reported in Ref.~\cite{Nalitov_PRL2017} but with anti-phase modulation of the real- and imaginary-parts of the lattice potentials (i.e., there regions of $\text{Re}{(V)}>0$ corresponded to $\text{Im}{(V)}<0$).
\begin{figure}[b!]
	\centering
	\includegraphics[width=0.8\columnwidth]{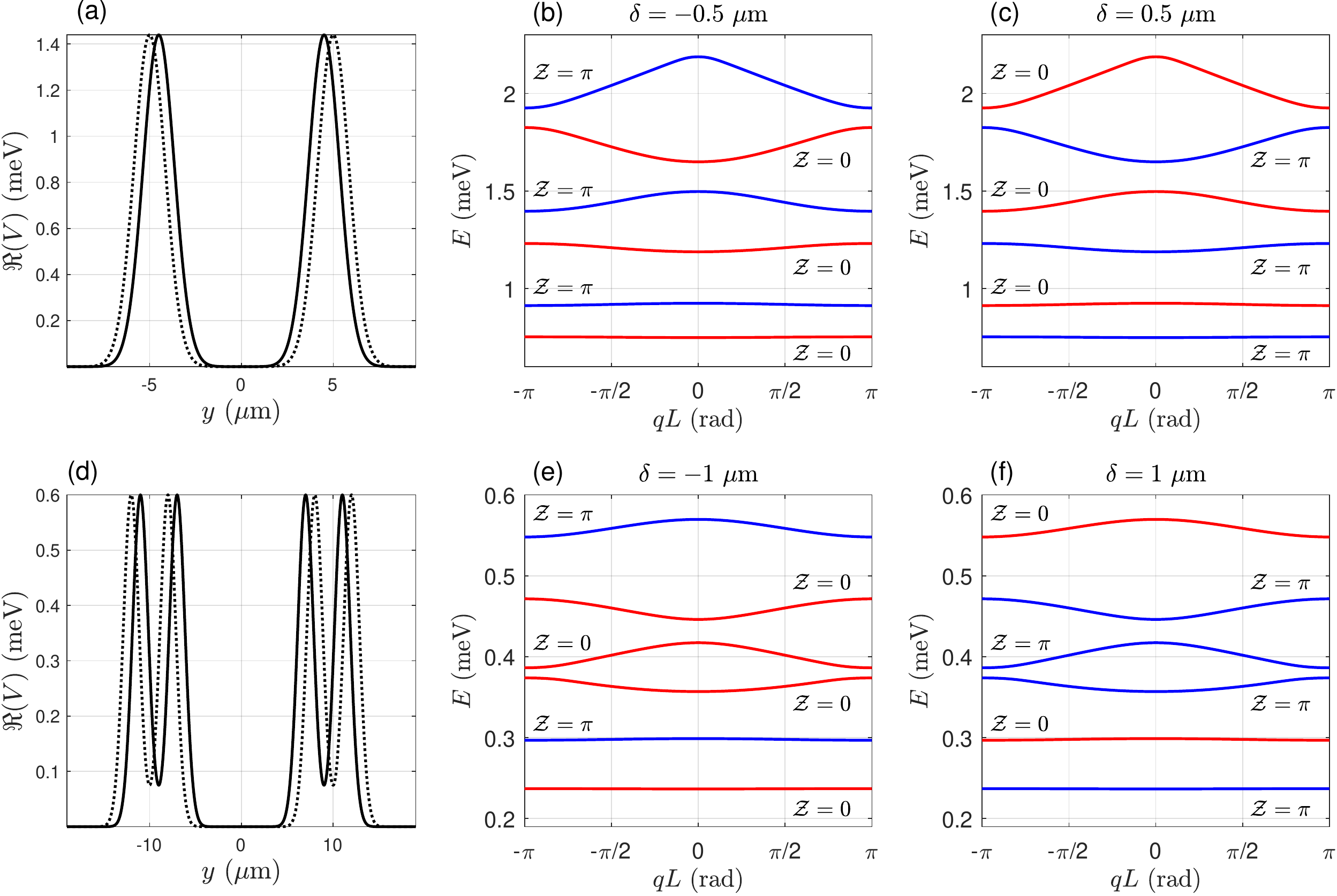}
	\caption{(a,d) Unit cell of the distance staggered (dimerised) potentials under consideration for calculation of Bloch states and their Zak phases. Solid and dotted lines denote the two choices of origin with inversion symmetry. Panels (b,c) show band structure and Zak phase of each band corresponding to potential (a) for the two distinct inversion symmetry points of the lattice. Panels (b,c) show the same calculation but for the potential given in panel (d).}
	\label{fig.blochcont}
\end{figure}

\subsection{First principle calculation of the Zak phase}
The continuum Bloch model [Eq.~\eqref{eq.bloch_eig}] can be applied to calculate the geometric phase arising in periodic systems. A well known formula for the geometric (Berry) phase accumulated when the wavefunction traverses a closed path in parameter space is written~\cite{Wilczek_Geometric_1989},
\begin{equation}
\gamma_C = \oint_C \langle u_{n,q} | \partial_q |u_{n,q} \rangle \, dq,
\end{equation}
where the integral is taken over a closed path $C$ in the Bloch states parameter space, typically the Brillouin zone's edge. For a one-dimensional (1D) lattice system the geometric phase $\gamma_C$ is called the {\it Zak phase}~\cite{Zak_PRL2989}, which we label $\mathcal{Z}$. We will be concerned with the discretisation of the Brillouin zone in small reasonable steps of momentum $q \in \{0, \delta_q, 2\delta_q, \dots J \delta_q \}$ where $J$ is an integer, and $J \delta_q = 2\pi/L$, and $L$ is the size of the lattice unit cell. We can write a more transparent form to our discretised crystal momentum parameter $q_j = j \delta_q$. A convenient gauge invariant method to numerically calculate the Zak phase of the $n$-th band takes advantage of the fact that $u_{n,q+\delta_q} \approx u_{n,q} + \partial_q u_{n,q}dq$, and the accumulated phase can then be written,
\begin{equation}
\mathcal{Z} = \text{Im} \log{ \prod_{j=0}^{J-1} } \langle u_{m,q_j} |u_{m,q_{j+1}} \rangle,
\end{equation}
where it is understood that $u_{m,q_N}(y) = e^{-i G y} u_{m,q_0}(y)$ (see e.g., Refs.~[\onlinecite{Resta_RevModPhys1994}] and [\onlinecite{Vanderbilt_PRB1993}]). Results are presented in Fig.~\ref{fig.blochcont} for two separate cases. In Fig.~\ref{fig.blochcont}(a,b,c) we scrutinise the case of distance staggered complex Gaussian potentials where the unit cell of the system is written $L =a_l+a_s =  (d+\delta) + (d-\delta) = 2d = 20$ $\upmu$m. For $\delta = 0.5$ $\upmu$m we observe each band interchangebly swapping from $\mathcal{Z} = 0$ and $\pi$. When $\delta = -0.5$ $\upmu$m the dimerisation is reversed and all bands undergo a $\pi$ change in their Zak phases, marking a topologically phase transition similar to reports in photonic crystal systems~\cite{Xiao_PRX2014}. Here we set $V_0 = 1.44 +i0.5$ meV.

In Fig.~\ref{fig.blochcont}(d,e,f) we consider a different geometry which can be accessed with current experimental technology. Consider a pair of Gaussians brought closely together forming a tight confining potential like a tuning fork. Next, consider two pairs of such Gaussian forks which allow polaritons in one fork to interfere with polaritons in the next (see Fig.~\ref{fig.blochcont}d). When the distance between the centres of the forks is dimerised the resulting Zak phases [Fig.~\ref{fig.blochcont}(e,f)] now display a different order from that of Fig.~\ref{fig.blochcont}(b,c). Indeed, the third and the fourth band now form a pair sharing the same Zak phase value. When the dimerisation is reversed, the Zak phase of each band changes by a unit of $\pi$ marking a topological phase transition. Here we set $V_0 = 0.6 +i0.1$ meV.

\subsection{Complex Su-Schrieffer-Heeger model of gain-localised polaritons}
If the system of weakly interacting condensates is discretised (see Section~\ref{sec.TB}) then it can be shown that the staggered interference between condensates makes up a non-Hermitian version of the Su-Schrieffer-Heeger (SSH) model~\cite{Su_PRL1979}. That is, when the chain of pumps are distance staggered so the system is described with a short distance $a_s = d - \delta$ and long distance $a_l = d+\delta$ (dimerised system) then the exchange of energy between polaritons at neighboring sites is described by two non-Hermitian hopping amplitudes $J_{\pm}$,
\begin{equation} \notag \label{eq.J}
J_\pm =  \eta \left(V_0\cos{(k_c [d\pm\delta])} - \frac{\hbar^2k}{\mu} \sin{(k_c [d\pm\delta])} \right)  |H_0^{(1)}(k_c[d\pm\delta])|.
\end{equation}
We will focus first on the bulk properties of such a dimerised system where we define $|n,\alpha \rangle$ as the polariton state in unit cell $n$ on sublattice $\alpha$ in a chain of $N$ pumps with periodic boundary conditions ($|1\rangle = |N+1\rangle$). The periodic boundary conditions are readily reproduced in a polariton experiment using a ring-shaped chain of pump spots such as shown in Fig.~1 in the main text (and with simulations shown in Fig.~\ref{fig.octagon}). The Hamiltonian describing the single-particle polariton chain system then reads,
\begin{equation} \label{eq.ssh}
\mathcal{H} = \Omega \sum_{n=1}^N \sum_\alpha |n,\alpha \rangle \langle n,\alpha| +  J_+ \sum_{n=1}^N(|n,B\rangle \langle n,A| + \text{h.c.})  + J_- \sum_{n=1}^{N-1} (|n+1,A\rangle \langle n,B| + \text{h.c.} ).  
\end{equation}
This can be diagonalised in a transparent manner by transforming to the basis of crystal momentum $|q\rangle = N^{-1/2} \sum_{n=1}^N e^{inq} |n\rangle$,
\begin{equation} \label{eq.Hq}
\mathcal{H}_q = 
\langle q | \mathcal{H} | q \rangle = \begin{pmatrix} \Omega & J_- + J_+ e^{iq} \\ J_- + J_+ e^{-iq} & \Omega
\end{pmatrix}, \qquad q \in \{\delta_q, 2\delta_q ,3\delta_q, \dots \} 
\end{equation}
where $\delta_q = 2\pi/N$. The Bloch solutions belonging to the eigenproblem of Eq.~\eqref{eq.Hq} are,
\begin{equation}
|b^{(\pm)}\rangle = \frac{1}{\sqrt{2}} \begin{pmatrix}
\pm 1 \\
e^{i \phi(q)} 
\end{pmatrix},
\end{equation}
where $(\pm)$ denotes the upper (conduction) and lower (valence) band of the system. The Zak phase~\cite{Zak_PRL2989}, which can only take values $0$ or $\pi$ (modulo $2\pi$) when the origin is chosen at an inversion center of the system, can be calculated straightforwardly by integration over the Brillouin zone,
\begin{equation} \label{eq.Zak_q}
\mathcal{Z} = -\frac{1}{2} \int_\text{BZ} \frac{\partial \phi(q)}{\partial q} \, dq.
\end{equation}
In Fig.~4  in the main text we give results on a band structure (red curves) fitted to experiment and calculated by diagonalising Eq.~\eqref{eq.Hq}. The parameters are: $V_0 = 1.44 + i0.5$,  $k_c = 1.5$ $\upmu$m$^{-1}$, $\Omega = 1.315$ meV, $d = 9.5$ $\upmu$m, $\delta = 0.5$ $\upmu$m, $m = 0.32$ meV ps$^2$ $\upmu$m$^{-2}$, and $\eta = 0.24$. The corresponding Zak phases shown in Fig.~4(b,c) [main text] are calculated using Eq.~\eqref{eq.Zak_q}.
\begin{figure}[b!]
	\centering
	\includegraphics[width=0.8\columnwidth]{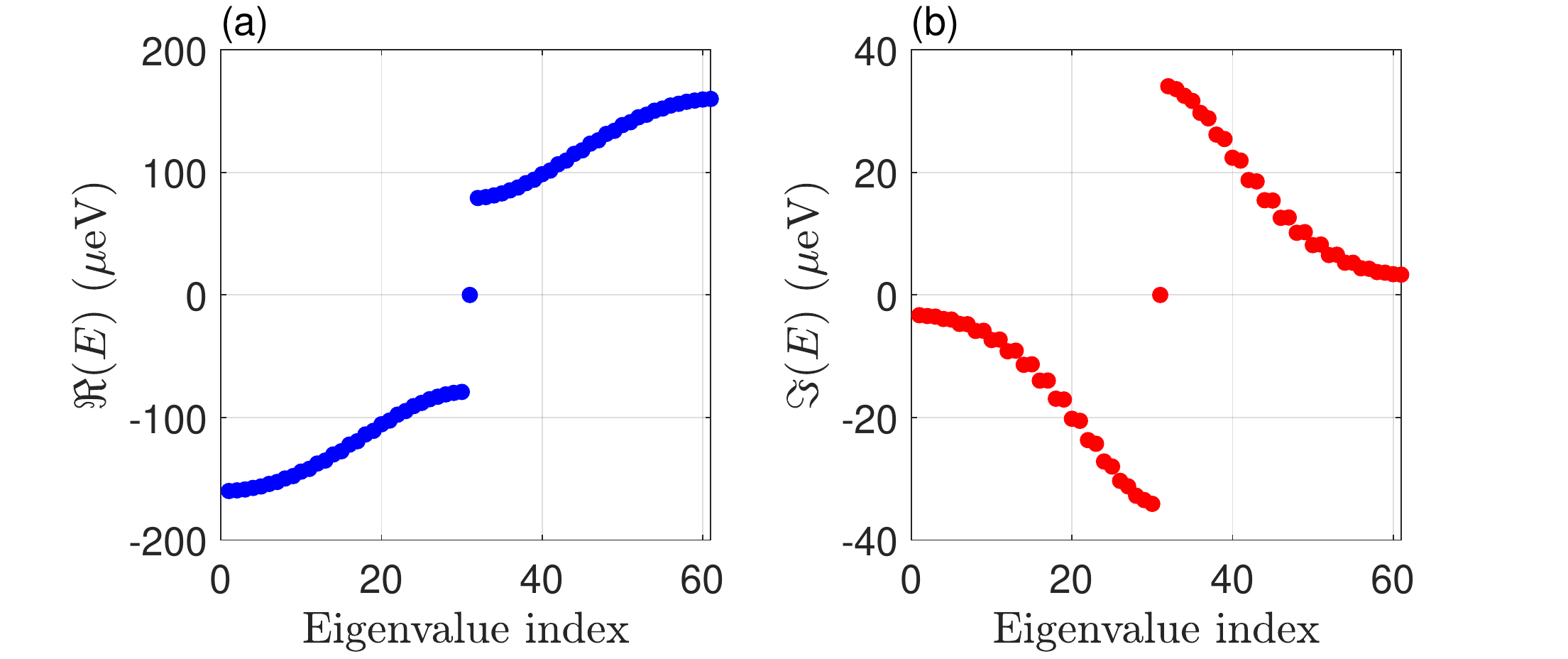}
	\caption{(a) Real and (b) imaginary parts of energies for a periodic system of $M=61$ sites which includes a defect linked to both its neighbours with a $J_+$ hopping amplitude. The results show a defect state lying at zero energy.}
	\label{fig.ssh}
\end{figure}

Moreover, we can optically engineer a defect state in the middle of the lattice, mimicking the behaviour of solitons in the polyacetylene polymers of the original SSH model~\cite{Su_PRL1979, Heeger_RMP1988}. Such a defected system, depicted as $\dots$ -A-B-A-B-A-A-B-A-B-A-B $\dots$, can be emulated by exciting a dimerised circle of polariton condensates, such as illustrated in Fig.~1d in the main text and Fig.~\ref{fig.octagon}c, and having one condensate adjacent to either two short-distance or two long-distance neighbours. The generation of the SSH dimerisation and defect states here is analogous to the engineering of a controllable phase factor (Peierls substitution) for the hopping amplitudes between adjacent sites in cold-atom systems using laser-assisted tunneling~\cite{Ruostekoski_PRA2008}. Setting $\Omega = 0$ for brevity, we plot in Fig.~\ref{fig.ssh} the real and imaginary parts of the eigenvalues for a periodic system of $61$ sites (i.e., $N = 30 + 1/2$ unit cells) in the basis of $|n,\alpha \rangle$ with a defect site linked by $J_+$ hoppings at each neighbour. The defect (midgap) state can be clearly distinguished from the rest as it lies on $E = 0$. The parameters used are the same as given below Eq.~\eqref{eq.Zak_q}. We point out that different from the topologically nontrivial defect states discussed here, in Section~\ref{sec.defect} we show standard defect condensation into midgap states between different orbital bands.

We point out that the exact details of engineering the dimerised lattice are not relevant in the long-wavelength limit (i.e., wavelengths larger than the distance staggering $\delta$) where interesting physical effects can arise. For instance, it is known that the bands at the edges of the reduced Brillouin zone can be linearised and described, in the continuum limit, with a relativistic Dirac Hamiltonian~\cite{Takayama_PRB1980, Ruostekoski_PRA2008}. The same limit can be argued for polaritons with Bloch momenta $q=\pi/2d$ by discretising the kinetic operator in steps appropriately (asymmetrically) coinciding with the lattice sites $n$, 
\begin{equation}
\partial_x^2 \sum_n \psi_n(x) \to \sum_n \frac{\psi(n-1) - 2 \psi(n) + \psi(n+1)}{(d-\delta)(d+\delta)}.
\end{equation}
It is understood that for a wave traveling along the chain with momentum $q=\pi/2d$ one has $\psi(n+1) = \psi(n) e^{i \pi/4}$, and therefore $\psi_{n-1} = - \psi_{n+1}$. 
\begin{equation}
\sum_n \frac{\psi(n-1) - 2 \psi(n) + \psi(n+1)}{(d-\delta)(d+\delta)} = -2 \sum_n \frac{\psi(n)}{d^2 - \delta^2} \approx -2 \sum_n \frac{\psi(n)}{d^2}.
\end{equation}
For the first order derivative one has,
\begin{equation}
\partial_x \sum_n \psi_n(x) \to \sum_n \frac{\psi(n+1) - \psi(n-1)}{2d} = \sum_n \frac{\psi(n + 1)}{d}.
\end{equation}
The action of the second order derivative can therefore be approximately interpreted as $\partial_x^2 \psi(x) \approx -2i \partial_x \psi(x)/d$ which, in the spinor representation of the two sublattices, is the same result shown in Refs.~\cite{Takayama_PRB1980, Ruostekoski_PRA2008}. More importantly, by engineering a defect in the lattice (such as shown in Fig.~\ref{fig.octagon}c) we create a kink in the condensate, permitting study of a polaritonic analogue of the Jackiw-Rebbi model. Originally, the model describes a 1D Dirac field coupled to a soliton background field containing a phase kink, resulting in topologically protected edge states and peculiar physics such as a fractional fermion number~\cite{Jackiw_PRD1976}. In the continuum limit (large lattice) the long wavelength physics of the condensatets can then be described by a (1+1)D relativistic Dirac equation for a polariton sublattice spinor $\chi = (\psi_A,-\psi_B)^T e^{i\pi/4}$, where $\psi_A$ and $\psi_B$ denote polaritons at each sublattice of the dimerised chain, 
\begin{equation}
H_D = \int  \left(-i c\hbar \chi^\dagger \hat{\sigma}_y \frac{\partial}{\partial x} \chi + \hbar g \varphi \chi^\dagger \hat{\sigma}_x \chi \right) \, dx.
\end{equation}
Here $c = d(J_+ + J_-)$, $g = J_+ - J_-$, and $\varphi$ denotes the soliton profile through the engineered defect in the condensate lattice.

\section{Tight binding treatment of gain localised polaritons} \label{sec.TB}
In this section we derive a single band picture of the weakly interacting condensates in the non-Hermitian lattice of Gaussian potentials. Our starting point will be Eq.~\eqref{eq.schro} considering the interaction between two polariton wavefunctions, gain-localised at their respective potentials, and separated by a distance $|y_1-y_2| = d$. Since the band physics apply only to the longitudinal axis of the lattice we will treat our problem as a 1D one. The basis of two gain-localised wavefunctions is chosen as,
\begin{equation} \label{Eq.Superposition}
\Psi(y,t) = c_1(t)  \phi_{1}(y) + c_2(t) \phi_{2}(y),
\end{equation}
where
\begin{equation}
\phi_n(y) = \sqrt{\kappa}e^{i k|y-y_n|}, \qquad k = k_c + i \kappa.
\end{equation}
Here $k_c,\kappa >0$ represents the outflow momentum and decaying envelope of the polaritons generated at each potential. Away from the pump spot the decaying polaritons will adopt an envelope given by $\kappa = m \gamma/2\hbar k_c$~\cite{Wouters_2008PRB}. For typical values from experiment, such as $k_c = 1.7$ $\upmu$m$^{-1}$, we have $\kappa = m \gamma/2\hbar k_c =  0.026$ $\upmu$m$^{-1}$, where  $m = 0.32$ meV ps$^2$  $\upmu$m$^{-2}$ and $\gamma = 1/5.5$ ps$^{-1}$. The rate of change in the wavefunction as a function of the spatial coordinate is then dominated by its momentum $k_c$, and we can drop $\kappa$ where appropriate in the following analysis. The small width of the potentials (2 $\upmu$m FWHM) allows us to treat the envelope of the wavefunction $\Psi(y)$ as slowly varying with the potentials approximately regarded as delta-like scatterers when $2\pi/k_c >$ FWHM.

Taking the inner product over the coordinate basis we have several separate integrals to evaluate in order to find the discretised set of equations,
\begin{equation}
i \hbar \int \Psi(y,t)^* \frac{d \Psi(y,t)}{dt} \, dy= \int \Psi(y,t)^* \left( -\frac{\hbar^2\partial_y^2}{2m} + V_0 \sum_{n=1}^{N} \delta{(y - y_n)} - i \frac{\hbar \gamma}{2} \right) \Psi(y,t) \, dy
\end{equation}
The integrals which arise can be listed as follows:
\begin{equation} \notag
\int_{-\infty}^{\infty} \phi_1^*  \left( - \frac{\hbar^2 \partial_y^2}{2m} \phi_1 \right) dy = \frac{\hbar^2}{2m}\left[ k^2  - 2ik \kappa   \right] 
\end{equation}
\begin{align} \notag
\int_{-\infty}^{\infty} \phi_2^*  \left( - \frac{\hbar^2 \partial_y^2}{2m} \phi_1 \right) dy
& = - \frac{\hbar^2 }{2m}\left[ 2 i k \kappa e^{-ik_c d} -k^2 \left( \cos(k_cd) +\frac{\kappa}{k_c} \sin(k_c d)   \right)    \right] e^{-\kappa d}  \\
& \approx - \frac{\hbar^2 }{2m}\left[ 2 i k \kappa e^{-ik_c d} -k^2  \cos(k_cd)  \right] e^{-\kappa d}
\end{align}
The approximation is valid for our polariton systems since $\kappa \ll k_c$. The rest of the integrals become,
\begin{align}
\int_{-\infty}^{\infty} \phi_1^* V(y) \phi_1 dy  & = V_0 \int_{-\infty}^{\infty} |\phi_1|^2 (\delta(y) + \delta(y - d)) dy   = V_0 \kappa(1 + e^{-2 \kappa d}), \\
\int_{-\infty}^{\infty} \phi_2^* V(y) \phi_1 dy  & = V_0 \int_{-\infty}^{\infty} \phi_2^* \phi_1 (\delta(y) + \delta(y - d)) dy  =2 V_0 \kappa e^{-\kappa d} \cos{(k_c d)},\\
\int_{-\infty}^{\infty} \phi_2^*\phi_1 dx  & \approx e^{-\kappa d}\cos{(k_c d)}.
\end{align}
Equation~\eqref{eq.schro} is now in the form of an algebraic differential equation,
\begin{equation} \label{eq.c}
i \hbar  \frac{dc_{1,2}}{dt}   + i \hbar e^{-\kappa d}\cos{(k_c d)}   \frac{dc_{2,1}}{dt}   =   F(c_{1,2},c_{2,1}) 
\end{equation}
where
\begin{align} \notag
F(c_1,c_2)  & =  \left( \frac{\hbar^2 k^2}{2m} - i\frac{\hbar\gamma_c}{2} \right) \left(c_1 + e^{-\kappa d}\cos{(k_c d)}  c_2 \right)   +  \left( \left[  V_0 - \frac{i\hbar^2k}{m}  \right] c_1+ V_0 e^{ikd} c_2 \right) \kappa  \\
& +  \left( \left[  V_0 - \frac{i\hbar^2k}{m}  \right] c_2 + V_0 e^{ikd} c_1 \right) \kappa e^{-ik^* d}.
\end{align}
Since $e^{-\kappa d}\cos{(k_c d)}<1$ for $0<d$ the differential-algebraic system of equations can be written as a system of ordinary differential equations. Additionally, assuming the coupling between the wavefunctions is weak, i.e. small $\xi = e^{- \kappa d}$ such that all terms of order $\mathcal{O}(\xi^2)$ and higher can be omitted, we arrive at an expression describing the interference between two spatially separated, gain localized, polariton wavefunctions,
\begin{equation}
i \hbar \frac{dc_{1,2}}{dt} =  \left[ \frac{\hbar^2 (k_c^2 + \kappa^2)}{2m} - i\frac{\hbar\gamma}{2} + \kappa V_0   \right] c_{1,2} +\kappa \left(V_0\cos{(k_c d)} - \frac{\hbar^2(k_c+i\kappa)}{m} \sin{(k_c d)} \right) e^{-\kappa d} c_{2,1}.
\end{equation}
Assuming that the momenta $k_c$ and the envelopes $\kappa$ of the gain-localised polariton wavefunctions are weakly affected by the coupling, the natural frequencies of each condensate is constant, and the correction in energy for the weakly interacting condensates is given by the following normal-mode splitting problem,
\begin{equation}
i \hbar \frac{dc_{1,2}}{dt}  \approx  \Omega c_{1,2} +\kappa \left(V_0\cos{(k_c d)} - \frac{\hbar^2k_c}{m} \sin{(k_c d)} \right) e^{-\kappa d} c_{2,1}.
\end{equation}
Here $\Omega$ is a constant denoting the natural frequency of the decoupled wavefunctions. In order to capture the correct distance dependence of the envelope we must replace the exponential term $\kappa e^{-\kappa d}$ from the 1D treatment with the 2D envelope of the Hankel function~\cite{Wouters_2008PRB},
\begin{equation}
i \hbar \frac{dc_{1,2}}{dt} =  \Omega c_{1,2} +\eta \left(V_0\cos{(k_c d)} - \frac{\hbar^2k_c}{m} \sin{(k_c d)} \right) |H_0^{(1)}(k_cd)| c_{2,1}.
\end{equation}
where $\eta$ is a fitting parameter. In the long wavelength limit where $k_c d\ll 1$ and the Hankel function is replaced by some suitably smooth function, the model assumes the standard tight binding form with $\cos{(k_c d)} \approx 1$, and $\sin{(k_c d)} \approx 0$. Extending the above analysis to a system of multiple coupled gain-localized wavefunctions in a chain we can write,
\begin{equation} \label{eq.schro_disc}
\text{i}\hbar \frac{dc_i}{dt}  =  \Omega c_i  +  \sum_{\langle ij \rangle} J(k_c|y_i-y_j|) c_{j},
\end{equation}
where the sum goes over nearest neighbours and the function $J$ captures the coupling,
\begin{equation} \label{eq.J}
J(k_c|y_i-y_j|) =  \eta \left(V_0\cos{(k_c|y_i-y_j|)} - \frac{\hbar^2k_c}{m} \sin{(k_c|y_i-y_j|)} \right)  |H_0^{(1)}(k_c|y_i-y_j|)|.
\end{equation}
We note that in the continuum limit the above discretised treatment describes a single polariton crystal band for $|y_n-y_m| =$ const. Equation~\eqref{eq.schro_disc} then stays accurate if the polaritons generated by the pumps occupy only a single band well separated from all other bands in the 1D pump lattice. It is then understood that the dimerisation of the lattice and consequent splitting of the aforementioned occupied band can be described using Eq.~\eqref{eq.schro_disc} when the dimerisation is small (i.e., small changes in $|y_n-y_m| = d \pm \delta$ where $\delta \ll d$).

\section{Simulations of defect condensation} \label{sec.defect}

\begin{figure}
	\centering
	\includegraphics[width=0.8\columnwidth]{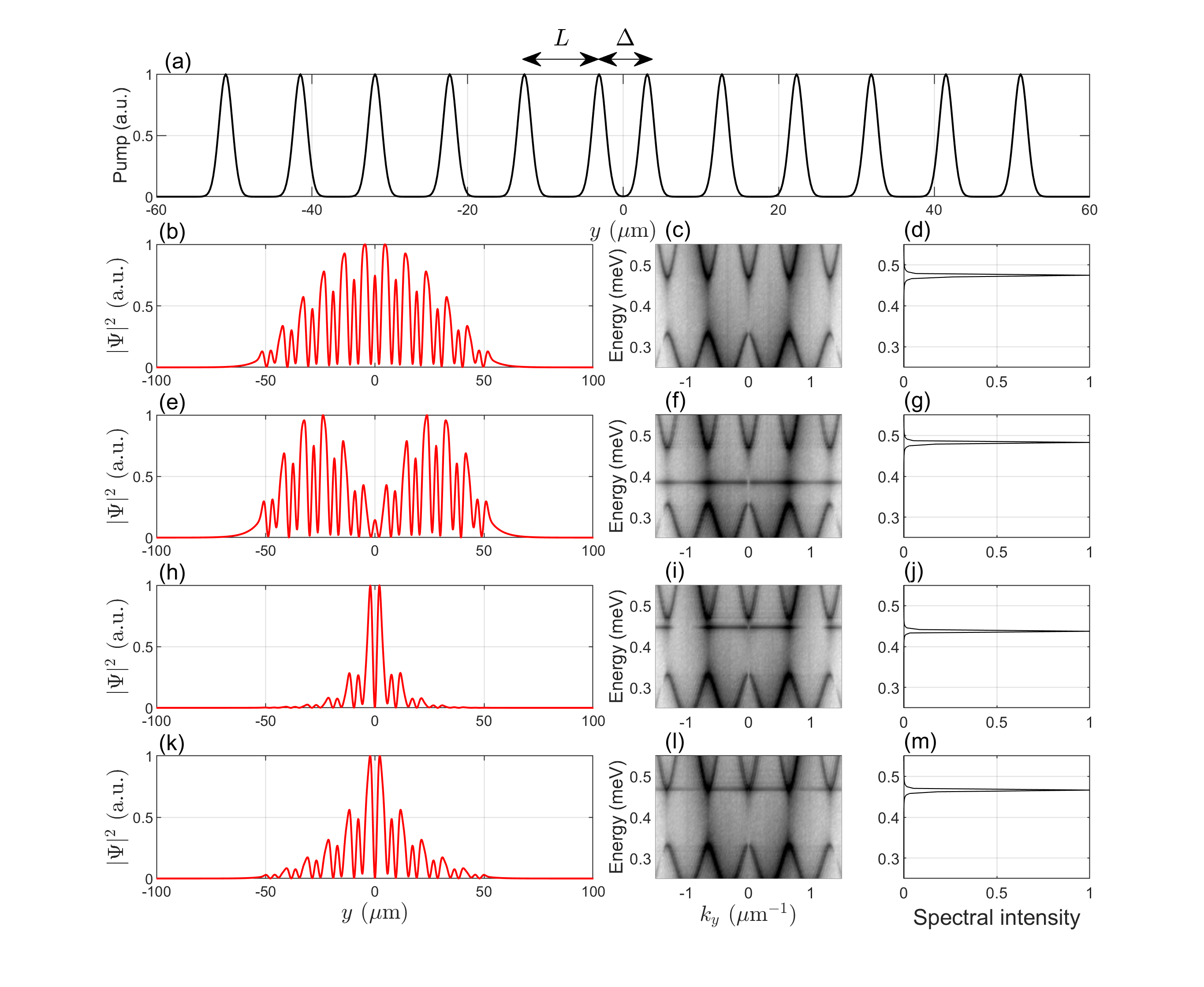}
	\caption{(a) Pump profile $P(y)$ plotted as a function of the lattice longitudinal coordinate illustrating. Here we choose 12 pumps with a lattice constant $L = 9.6$ $\upmu$m and a defect in the center denoted by distance $\Delta$. Panels (b,e,h,k) show the numerically simulated 1D steady state condensate intensity in real space and corresponding spectral output in (d,g,j,m). Here we have set $\Delta = 9.6,\, 8.0, \,7.1, \, 6.2$ $\upmu$m respectively. Panels (c,f,i,l) show corresponding numerically resolved single particle dispersion for clarity with defect state appearing in the bandgap.}
	\label{sim3Fig}
\end{figure}
If the translational symmetry of the lattice is broken by introducing a defect into the pump geometry gap states appear between bands belonging to different polariton orbitals/modes. The defect geometry is then found to either favor or disfavor condensation at its location  (see Fig.~5 in main text and Fig.~\ref{sim3Fig}). To illustrate this finding we simulate the 1D dGPE which is the same as Eq.~\eqref{eq.GP_Res} with the exception of,

\begin{equation}
\nabla^2 \to (1 - i \Lambda) \partial_y^2,
\end{equation}

where $\Lambda$ is a phenomenological energy damping parameter. The defected scenario is illustrated in Fig.~\ref{sim3Fig}a where $L = 9.6$ $\upmu$m denotes the lattice constant and $\Delta$ the size of the defect. Parameters are taken same as in Section~\ref{sec1} with the exception of $\hbar R = 17.8$ $\upmu$eV $\upmu$m$^{-1}$, $\Lambda = 0.1$, and $P_0 = 6$ $\upmu$m$^{-1}$ ps$^{-1}$.

In Fig.~\ref{sim3Fig}b we show the steady state condensate density in the finite optical polariton crystal for $\Delta/L=1$. The finite length of the crystal corresponds to open boundaries causing increased losses around the edges. In fact, the additional losses at the crystal edges can be regarded as an effective trapping potential due to the flux of particles from the system~\cite{Ostrovskaya_PRL2013}. In Fig.~\ref{sim3Fig}e we show the condensate steady state in a crystal with a defect at $\Delta/L= 0.83$ $\upmu$m corresponding to a flat defect/midgap state lifted from the lower band (see dark horizontal line Fig.~\ref{sim3Fig}f). Interestingly, the presence of the midgap state causes a sharp drop in condensate density even though more power is technically being supplied through the lasers in the middle of the crystal. This seemingly counterintuitive observation is due to the weak evanescent coupling of the defect mode to the rest of the crystal resulting in poor energy flow from the high gain region of the bulk into the defect mode. In Fig.~\ref{sim3Fig}h the defect size is decreased to $\Delta/L = 0.74$ and the condensate energy (see Fig.~\ref{sim3Fig}j) redshifts slightly as it start occupying the defect state, consequently becoming highly localised. Lastly, when the defect mode becomes resonant with the upper band at $\Delta/L = 0.65$ (see Fig.~\ref{sim3Fig}l) the condensate in the defect mode couples with the rest of the crystal and becomes more delocalised (see Fig.~\ref{sim3Fig}k). We label these shape preserving nonlinear modes of the condensate in Figs.~\ref{sim3Fig}e,h as {\it dissipative dark}- and {\it bright gap solitons} respectively, in analogy with the bright gap solitons observed previously in polariton lattices~\cite{Tanese_NatComm2013}.
	
\bibliographystyle{naturemag}
\bibliography{refs}
\end{document}